# Magnetodynamic Characteristics and QGP Energy Dissipation in RMHD Framework with Relativistic Heavy-Ion Collisions

Huang-Jing Zheng[1] and Sheng-Qin Feng[1,2,3,*]

[1] College of Science, China Three Gorges University, Yichang 443002, China

[2]Center for Astronomy and Space Sciences and Institute of Modern Physics, China Three Gorges University, Yichang 443002, China

[3] Key Laboratory of Quark and Lepton Physics (MOE) and Institute of Particle Physics, Central China Normal University, Wuhan 430079, China

**Abstract**

Relativistic heavy-ion collisions generate ultra-strong magnetic fields that interact with the quark – gluon plasma (QGP), a key focus of high-energy physics research. This study investigates QGP energy density evolution under time-dependent magnetic fields within a (1 + 1)D relativistic magnetohydrodynamic (RMHD) framework integrated with Bjorken flow. Three magnetic field temporal evolution models (Type-1, Type-2, Type-3) are analyzed for two different equations of state: (1) $p = c_s^2 \varepsilon$ (simplified ultra-relativistic), and (2) $p = c_s^2 \varepsilon - 2MB$ (magnetized conformal), incorporating a temperature-dependent mag netic susceptibility derived from lattice QCD. Results show that stronger magnetic fields consistently suppress QGP energy density decay, with suppression magnitude dependent on the magnetic field's temporal profile. Ultra-relativistic fluids exhibit slowed energy decay due to magnetic pressure counteracting hydrodynamic expansion. In contrast, magnetized conformal fluids display faster energy dissipation under identical conditions, arising from the synergistic effect of enhanced magnetic fluid coupling, increased energy dissipation during interaction, and QGP's perfect fluid expansion at elevated temperatures. Temperature-dependent magnetic susceptibility reveals a transition from diamagnetic (confined phase) to paramagnetic (deconfined QGP phase) behavior, introducing a feedback mechanism that strengthens energy retention at higher temperatures. This work clarifies the interplay between magnetic field dynamics, QCD phase structure, and hydrodynamic expansion, providing key observational signatures for distinguishing fluid types in heavy-ion collisions and advancing realistic modeling of magnetized QGP.

*corresponding author: fengsq@ctgu.edu.cn

## I INTRODUCTION

Relativistic heavy-ion collision experiments at facilities like the Relativistic Heavy Ion Collider (RHIC) provide a unique laboratory to study the properties of quantum chromodynamic (QCD) matter under extreme conditions. In such collisions, the interaction of two highly Lorentz-contracted nuclei generates a hot, dense medium dominated by deconfined quarks and gluons—the quark-gluon plasma (QGP)[1,2]. This short-lived state exhibits near-perfect fluidity, enabling relativistic hydrodynamic models to successfully describe its collective expansion and thermalization dynamics.[3-8]. Notably, the interplay between hydrodynamic evolution and electromagnetic phenomena has emerged as a critical frontier in understanding QGP behavior, particularly in non-central collisions where ultra-strong magnetic fields ( $10^{18} \sim 10^{19}$ Gauss) are transiently generated[9-16].

These magnetic fields, though short-lived in vacuum [17], persist longer in the electri cally conductive QGP medium [18–21], creating opportunities to probe magnetohydrody namic effects. Key phenomena such as the Chiral Magnetic Effect (CME) [22,23], Chiral Separation Effect (CSE) [24], and their collective manifestation as Chiral Magnetic Waves (CMWs) [25,26] are theorized to induce charge-dependent azimuthal anisotropies in parti cle emission [27]. However, disentangling these signals from background collective flow remains experimentally challenging [28–30], necessitating precise theoretical modeling of the coupled evolution of the QGP and electromagnetic fields [31–38]. Relativistic mag netohydrodynamics (RMHD) offers a self-consistent framework to explore this coupling. While prior studies have examined simplified scenarios, the time-dependent evolution of magnetic fields particularly their dependence on proper time (τ) in the RHIC energy regime remains under explored in (1 + 1)D RMHD frameworks. This gap limits our ability to quan tify how magnetic field dissipation shapes energy density dynamics and thermalization processes in the QGP.

It is widely believed that the QGP produced in high-energy heavy-ion collisions be haves as an almost perfect fluid. Perfect hydrodynamics assumes zero viscosity, implying that fluid elements move without internal "friction". However, in more realistic scenarios, the QGP is a fluid with small dissipative properties (viscosity). Several transport coefficients characterize the internal "friction" of the fluid. For example, shear viscosity $\eta$ (dynamic vis cosity) measures the fluid's resistance to flow, bulk viscosity $\zeta$ (volume viscosity) measures the fluid's resistance to expansion, and thermal conductivity $\lambda$ measures the fluid's ability to conduct heat.

In this work, we address this challenge by integrating three distinct models of time dependent magnetic field evolution into a (1 + 1) D relativistic magnetohydrodynamic framework based on Bjorken flow [39]. These models (Type-1, Type-2 and Type-3) capture characteristic $\tau$-dependencies observed in RHIC energy collisions [40–43]. We system atically analyze their impact on energy dissipation in two fluid scenarios: (1) an ultra relativistic fluid with a simplified equation of state ( $p = c_s^2 \varepsilon$ ), and (2) a magnetized conformal fluid incorporating explicit magnetization effects. To enhance physical realism, we further introduce a temperature-dependent magnetic susceptibility ($\chi_m$ ($T$)) derived from lattice QCD calculations [44–49], which encodes the transition from diamagnetic hadronic matter ($\chi_m < 0$) to paramagnetic QGP ($\chi_m > 0$).

Our study aims to resolve two pivotal questions: (1). How do the temporal profiles of magnetic fields in the RHIC regime influence the energy density evolution of QGP? (2). What role does the phase structure of Quantum Chromodynamics (QCD), manifested through temperature ($T$), play in regulating electromagnetic response and energy dissi pation? By bridging first-principles QCD inputs with macroscopic RMHD simulations, this work advances the quantitative understanding of magnetized QGP dynamics. The results not only clarify the interplay between magnetic field decay and hydrodynamic expansion but also establish benchmarks for future studies incorporating dissipative effects and spatial inhomogeneities.

The paper is organized as follows: Section 2 outlines the RMHD formalism with magnetization. Section 3 adapts Bjorken flow to incorporate external magnetic fields. Section 4 compares energy density evolution across the three magnetic field models for both fluid types. Section 5 integrates lattice QCD-derived $\chi_{\mathrm{m}}(T)$ and a realistic equation of state. Section 6 summarizes key findings and discusses implications for heavy-ion physics.

## II Magnetohydrodynamics with Bulk Viscosity and Magnetization

We expand on previous studies[50-52] to consider RMHD with both bulk viscosity and non-zero magnetization. In this case, the energy-momentum tensor is expressed as follows

$$T^{\mu\nu} = \left(e + p + \Pi - MB + B^2\right)u^\mu u^\nu - \left(p - MB + \Pi + \frac{B^2}{2}\right)g^{\mu\nu} + \left(MB - B^2\right)b^\mu b^\nu \quad , \tag{1}$$

where $e$ denotes the energy density, $p$ denotes the pressure, $\Pi$ represents the bulk viscous pressure. $B$ and $M$ denote the magnetic field strength and the magnetization intensity in the local rest frame of the fluid, respectively, where $M = \chi_m B$, and $\chi_m$ is the magnetic susceptibility. In the original Israel-Stewart (IS) theory, the bulk viscous stress $\Pi$ is considered an independent dynamical variable[53-55] given by the following equation

$$\begin{aligned} \Pi = {} & \Pi_{NS} - \tau_\Pi \dot{\Pi} \\ & + \tau_{\Pi_q} q \cdot \dot{u} - \ell_{\Pi_q} \partial \cdot q - \zeta \hat{\delta}_0 \Pi \theta \\ & + \lambda_{\Pi_q} q \cdot \nabla \alpha + \lambda_{\Pi_\pi} \pi^{\mu\nu} \sigma_{\mu\nu} \end{aligned} \quad , \tag{2}$$

where $\zeta$ is the bulk viscosity, and $\tau_\Pi$ is the relaxation time for the bulk viscosity. To further simplify, we consider only the first-order case. The subscript $NS$ denotes the Navier-Stokes value, which can be written as where $\zeta$ is the bulk viscosity, and $\tau_\Pi$ is the relaxation time for the bulk viscosity. First-order (Navier-Stokes) viscous hydrodynamics suffers from acausal propagation and instabil ity in the relativistic regime. Such violations of causality and stability are resolved by second-order theories such as the Israel – Stewart formalism, which treat dissipative fluxes as dynamical variables with relaxation times. While essential for rigorous transport mod eling, the Israel – Stewart framework introduces additional complexity without altering the qualitative magnetic field effects studied here. We thus employ first-order theory as a baseline, noting that our primary conclusions regarding magnetic suppression of energy dissipation remain robust when extended to second-order formulations [56].

The subscript NS denotes the Navier – Stokes value, which can be written as

$$\Pi_{NS} = -\zeta\theta = -\zeta\partial_\mu u^\mu , \tag{3}$$

where $\zeta = bs$ , $s$ is the entropy density, $b$ is the parameterized bulk viscosity coefficient [57,58].

The projection of energy-momentum along the four-dimensional velocity vector $u^\nu$ corresponds to the conservation of fluid energy, and its form is as follows:

$$\begin{aligned} 0 &= u_\nu \partial_\mu T^{\mu\nu} \\ &= u^\alpha \partial_\alpha \varepsilon + B u^\alpha u_\alpha B + (\varepsilon + p + \Pi)\partial_\alpha u^\alpha - MB\partial_\alpha u^\alpha \\ &\quad + B^2 \partial_\alpha u^\alpha + MB u_\mu b^\nu \partial_\nu b^\mu - B^2 u_\mu b^\nu \partial_\nu b^\mu \end{aligned} , \tag{4}$$

where $u_\nu u^\nu = 1$, $u_\nu b^\nu = 0$, and $u_\nu \partial_\mu u^\nu = 0$ are taken. The Maxwell equation is given by

$$\frac{1}{2}(u^\alpha \partial_\alpha) B^2 + B^2 \partial_\alpha u^\alpha + B^2 b^\mu b^\nu \partial_\nu u_\mu = 0 . \tag{5}$$

Similarly, the projection of the energy-momentum along the direction orthogonal to the four-velocity $u^\mu$ corresponds to the momentum conservation, and its expression is

$$\begin{aligned} 0 &= \Delta_{\nu\alpha} \partial_\mu T^{\mu\nu} \\ &= \Delta_{\nu\alpha}(\varepsilon + p + \Pi - MB + B^2) u^\mu \partial_\mu u^\nu + \Delta_{\nu\alpha}(\varepsilon + p + \Pi - MB + B^2) u^\nu \partial_\mu u^\mu \\ &\quad - \Delta_{\nu\alpha} \partial^\nu (p - MB + \Pi + \frac{B^2}{2}) + \Delta_{\nu\alpha}(MB - B^2)\partial_\mu (b^\mu b^\nu) + \Delta_{\nu\alpha} b^\mu b^\nu \partial_\mu (MB - B^2) , \\ &= (\varepsilon + p + \Pi - MB + B^2) u^\mu \partial_\mu u_\alpha - \Delta_{\nu\alpha} \partial^\nu (p - MB + \Pi + \frac{B^2}{2}) \\ &\quad + \Delta_{\nu\alpha} \partial_\mu \left[ \left( MB - B^2 \right) b^\mu b^\nu \right] \end{aligned} \tag{6}$$

and the last term of Eq. (6) is given as

$$\Delta_{\nu\alpha} \partial_\mu \left[ \left( MB - B^2 \right) b^\mu b^\nu \right] = (\chi_m - 1) \left[ \partial_\mu \left( B^\mu B^\alpha \right) - u_\nu u^\alpha \partial_\mu (B^\mu B^\nu) \right] , \tag{7}$$

where the magnetization coefficient $\chi_m$ is constant. It is found that the last term of Eq. (6) vanishes. The simplified momentum conservation becomes

$$\begin{aligned} 0 &= \Delta_{\nu\alpha} \partial_\mu T^{\mu\nu} \\ &= (e + p + \Pi - MB + B^2) u^\mu \partial_\mu u_\alpha - \Delta_{\nu\alpha} \partial^\nu (p - MB + \Pi + \frac{B^2}{2}) \end{aligned} , \tag{8}$$

where one uses $\Delta_{\nu\alpha} u^\nu = 0$ .

## III. Bjorken flow

In the work, we consider a fluid undergoing Bjorken expansion. For longitudinally boost-invariant flow, Milne coordinates are more convenient than standard Cartesian coordinates $(t,x,y,z)$. Milne coordinates are $x^{\mu}=(\tau,x,y,\eta_s)$, which are natural choices for describing ultra-relativistic heavy-ion collision, where $\tau=\sqrt{t^2-z^2}$ is the proper time and $\eta_s=\tanh^{-1}(z/t)$ is the pseudorapidity.

The Bjorken flow[39] is uniform in the transverse direction and exhibits longitudinal boost-invariant. Its symmetry lies in the transverse plane (x - y plane), enhancing invariance along the longitudinal (or beam) direction, and also exhibiting reflection symmetry $(z\to -z)$. This implies that the flow profile $v^x=v^y=0$ and $v^z=z/t$ are symmetric in the transverse plane and invariant along the longitudinal direction. The fluid velocity can be taken as

$$u^{\mu}=\left(\frac{t}{\tau},0,0,\frac{z}{t}\right) \quad , \tag{9}$$

where $\tau=\sqrt{t^2-z^2}$ is the proper time. All macroscopic quantities depend only on the proper time $\tau$, and are independent of the spatial coordinates and the four-velocity $\eta=\frac{1}{2}\left[(t+z)/(t-z)\right]$.

Using Milne coordinates, the four-velocity simplifies to $u^{\mu}=(1,\vec{0})$, the spatial derivatives and the four-divergence are given by

$$\begin{aligned} u^{\mu}\partial_{\mu} &= \partial_{\tau} \\ \partial_{\alpha}u^{\alpha} &= \theta = \frac{1}{\tau} \end{aligned} , \tag{10}$$

In this work, we adopt the Bjorken flow as a baseline scenario to isolate the effects of magnetic field evolution on energy density dynamics. While more general flow profiles including accelerating solutions (e.g., $v = tanh(\lambda\eta)$ in Rindler coordinates) exist and may be important in the initial stages [59], the Bjorken flow provides a transparent framework for comparing different magnetic field models and equations of state. Investigation of accelerating flows is left for future work.

In addition to the direction provided by the fluid velocity, the magnetic field also selects another special direction in the system, which we associate with a spacelike unit vector $b^{\mu}=B^{\mu}/B$, where $B=\sqrt{-B^{\mu}B_{\mu}}$, it is normalized to $b^{\mu}b_{\mu}=-1$. In our work, we assume that the magnetic field is perpendicular to the reaction plane and is oriented along the y-axis in a fluid with infinite conductivity and non-zero magnetization $B^{\mu}=(0,0,B_y,0)$.

In the following work, the useful thermodynamic relations are reviewed. For a perfect fluid in thermodynamic equilibrium, we have

$$\varepsilon+p=Ts+\mu n\,, \tag{11}$$

where $n$ is the baryon number density and $\mu$ is the corresponding chemical potential.These follow the standard thermodynamic relations

$$d\varepsilon = Tds + \mu dn - MdB \ , \tag{12}$$

From these thermodynamic relations, we can derive the Gibbs-Duhem relation as

$$dp = sdT + nd\mu + MdB \ . \tag{13}$$

In ultra-relativistic heavy-ion collisions, the net baryon number density and chemical potential are nearly zero at mid-rapidity. Therefore, for simplicity, we consider the case of zero baryon chemical potential in our work.

## IV. Fluid Evolution under Three Different Magnetic Field Conditions

Within the framework of linear approximation, the effects of magnetization can be characterized by magnetic susceptibility. A wealth of theoretical studies have indicated that the medium shows diamagnetic behavior in the confined phase i.e., $\chi_m < 0$ , whereas it exhibits paramagnetic behavior in the deconfined quark-gluon plasma (QGP) phase,i.e., $\chi_m > 0$ [44,49,51,60]. In the present work, we exclusively focus on the behavior of the magnetic field in the deconfined phase. For the sake of simplicity, we assume that magnetic susceptibility $\chi_m$ remains constant throughout this section, which thus leads to the following form of the energy conservation equation:

$$\partial_\tau e + \frac{(e + p + \Pi - MB + B^2)}{\tau} + \frac{1}{2}\partial_\tau B^2 = 0 \quad , \tag{14}$$

since the magnetic field is non-zero only in the $y$ direction, there exists

$$-B^2 u_\mu b^\nu \partial_\nu b^\mu = 0 = MBu_u b^\nu \partial_\nu b^\mu \quad , \tag{15}$$

Inserting Eq. (15) into Eq. (14), one can obtain

$$\partial_\tau e + \frac{e + p + \Pi}{\tau} + \frac{1}{2}\chi_m \partial_\tau B^2 = 0 \quad . \tag{16}$$

The external magnetic field generated by relativistic heavy ion collisions is introduced into the framework of relativistic magnetohydrodynamics, discussing the variation of energy density with proper time. It is assumed that the magnetic field $B_y(\tau, \vec{x}) = B_0(\vec{x}) F_B(\tau_B, \tau)$ [23] is along the y direction in relativistic heavy-ion collisions and consider three different parameterization approaches commonly used in the literature for studying various magnetic field effects in relativistic heavy-ion collisions as:

$$\text{Type-1:} \quad F_B(\tau_B, \tau) = \frac{1}{1 + (\tau - \tau_0)^2 / \tau_B{}^2} \quad , \tag{17}$$

Type-2: $F_B(\tau_B, \tau) = \dfrac{1}{[1 + (\tau - \tau_0)^2 / \tau_B{}^2]^{3/2}}$ , (18)

and Type-3: $F_B(\tau_B, \tau) = e^{-|\tau - \tau_0|/\tau_B}$ . (19)

In all these parameterizations, $\tau_B$ is the fundamental magnetic field lifetime parameter that controls the rate at which the magnetic field decreases over time. However, it should be noted that due to differences in their functional forms, the evolution of the magnetic field varies slightly for the same $\tau_B$ value[23]. In this study, it is assumed that $\tau_B$ is comparable to $\tau_0$, meaning that the lifetime of the magnetic field is similar to the initial time of hydrodynamic evolution. This assumption implies that the magnetic field exists during the early stages of hydrodynamic evolution but gradually vanishes over time. The required lifetime $\tau_B$ decreases with increasing beam energy. For example, $\tau_B \sim 5$fm/c for $\sqrt{s_{NN}}$ =11.5 GeV, $\tau_B \sim 0.5$ fm/c for $\sqrt{s_{NN}}$ = 200 GeV. Parameter A = 92 for type-1, A = 125 for type-2 and A = 128 for type-3. An average over these three types of time dependence in a (perhaps naive) statistical way would suggest $\tau_B = A/\sqrt{s_{NN}}$ with $A = 115 \pm 16$ GeV $\times fm/c$ .

In all these parameterizations, $\tau_B$ is the fundamental magnetic field lifetime parameter that controls the rate at which the magnetic field decreases over time. However, it should be noted that due to differences in their functional forms, the evolution of the magnetic field varies slightly for the same $\tau_B$ value [23]. In this study, it is assumed that $\tau_B$ is comparable to $\tau_0$, meaning that the lifetime of the magnetic field is similar to the initial time of hydrodynamic evolution. This assumption implies that the magnetic field exists during the early stages of hydrodynamic evolution but gradually vanishes over time. To quantify the energy dependence of the extracted magnetic field lifetime, we fit the optimal $\tau_B$ values to a scaling formula motivated by the Lorentz contraction of the nuclear passing through time: $\tau_B = A/\sqrt{s_{NN}}$ , where the fitting parameter A (with units of GeV × fm/c) characterizes the overall magnitude of the magnetic field lifetime across collision energies. For example, $\tau_B \sim 5$fm/c for $\sqrt{s_{NN}}$ =11.5 GeV, $\tau_B \sim 0.5$ fm/c for $\sqrt{s_{NN}}$ = 200 GeV. Parameter A = 92 for type-1, A = 125 for type-2 and A = 128 for type-3. An average over these three types of time dependence in a (perhaps naive) statistical way would suggest $\tau_B = A/\sqrt{s_{NN}}$ with $A = 115 \pm 16$ GeV $\times fm/c$ .

Both fluids considered in this work are ultra-relativistic. The distinction lies in their treatment of magnetization and conformality: the first case is a simple ultra-relativistic ideal fluid with $p = c_s^2 \varepsilon$ and no explicit magnetization correction. The second case is a magnetized conformal fluid, where the term -2*MB* is included to maintain conformality at finite magnetic field *B*. Conformality is strictly satisfied only for $c_s = 1/\sqrt{3}$ .

### A. Ultra-Relativistic Ideal Fluid (Without Magnetization Correction)

Since the fluid is considered as ultra-relativistic, the contribution of the rest mass to the equation of state (EOS) can be neglected. The pressure is proportional to the energy density, i.e., $p = c_s{}^2 e$ , where $c_s$ is the local sound speed, which is assumed to be constant ($c_s = 1/\sqrt{3}$). Some dimensionless quantities $\tilde{e} = e/e_0$ , $\sigma_0 = B_0{}^2/e_0$ , $\tilde{p} = p/e_0$ are defined to describe the energy density, and the pressure varies with proper time, where, $e_0$ represents the initial energy density.

Substituting the magnetic field decay models under the three different modes into Eq., we obtain the following form:

$$\text{Type-1:} \quad \partial_\tau \tilde{e} + (1 + c_s{}^2)(\frac{\tilde{e}}{\tau} - b\frac{\tilde{e}}{\tau^2 T}) - \frac{2\chi_m(\tau - \tau_0)\sigma_0}{[1 + (\tau - \tau_0)^2/\tau_B{}^2]^3 \tau_B{}^2} = 0 \quad , \tag{20}$$

$$\text{Type-2:} \quad \partial_\tau \tilde{e} + (1 + c_s{}^2)(\frac{\tilde{e}}{\tau} - b\frac{\tilde{e}}{\tau^2 T}) - \frac{3\chi_m(\tau - \tau_0)\tau_B{}^6\sigma_0}{(\tau^2 - 2\tau\tau_0 + \tau_0{}^2 + \tau_B{}^2)^4} = 0 \ , \tag{21}$$

$$\text{Type-3:} \quad \partial_\tau \tilde{e} + (1 + c_s{}^2)(\frac{\tilde{e}}{\tau} - b\frac{\tilde{e}}{\tau^2 T}) - \frac{\chi_m \sigma_0 e^{-\frac{2|\tau - \tau_0|}{\tau_B}}}{\tau_B}\frac{\partial|\tau - \tau_0|}{\partial_\tau} = 0 \quad , \tag{22}$$

Figure 1 investigates the evolution of energy density in an ultra-relativistic fluid under the influence of time-dependent magnetic fields in the relativistic heavy-ion collisions within the framework of relativistic magnetohydrodynamics (RMHD). The figure depicts three different magnetic field evolution models (Type-1, Type-2, and Type-3) while considering various magnetic field values of $\sigma_0$. Specifically, it examines how three distinct magnetic field evolution models (Type-1, Type-2, and Type-3 in relativistic heavy-ion collisions) affect the decay rate of energy density as a function of proper time ($\tau$). The study highlights the interplay between magnetic field strength (parameterized by $\sigma_0$) and energy dissipation dynamics. All the models indicate that, compared with the situation without magnetic field influence, the initial stronger magnetic field leads to a slower energy density decay.

Figure 1 underscores the critical role of magnetic field evolution models in shaping energy density dynamics. The study bridges theoretical RMHD frameworks with relativistic heavy-ion experimental, emphasizing the need for precise modeling of magnetic fields in high-energy physics.

We note that the present framework includes bulk viscosity through the Israel–Stewart formalism (Equations (2) and (3)), with the Navier–Stokes value $\Pi_{NS} = -\zeta\theta$. While bulk viscosity alone can produce a 'reheating' effect that slows energy density decay [59], here we focus on isolating the magnetic field contribution by fixing the bulk viscosity coefficient. The interplay between magnetic suppression and viscous dissipation - particularly whether these effects amplify or compete with each other—presents an interesting direction for future systematic studies with varying $\zeta$.

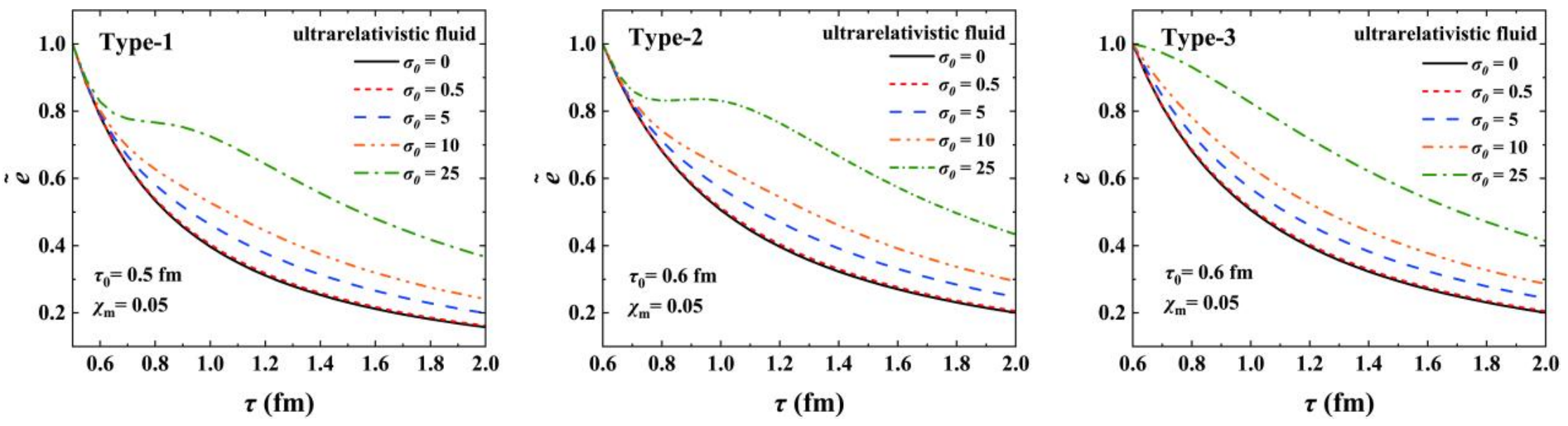

**Figure 1.** Evolution of energy density in the ultra-relativistic fluid. We choose magnetic susceptibility $\chi_m = 0.05$ , $b = 0.1$ , $T = 200$ MeV [57].The left, middle, and right panels correspond to three different temporal evolution models of the magnetic field. The black solid line represents the case without a magnetic field. The red short dashed line, blue long dashed line, orange dash-dot-dotted line, and green dash-dotted line correspond to different values of $\sigma_0 = 0.5, 5, 10, 25$ , respectively.

## B. Magnetized Conformal Fluid (With Conformality-Preserving Trace Correction)

In this case, the energy density takes into account the degree of magnetization or the strength of the magnetic field, i.e., a conformal fluid with a magnetic field present in four-dimensional spacetime

$$e = \frac{1}{c_s^2} p - 2MB \ . \qquad (23)$$

Then, the conservation equation under different magnetic field models becomes

Type-1： $$\partial_\tau \tilde{e} + (1+c_s^2)\left(\frac{\tilde{e}}{\tau} - b\frac{\tilde{e}}{\tau^2 T}\right) + \frac{2\chi_m \sigma_0}{[1+(\tau-\tau_0)^2/\tau_B^2]^2}\left(\frac{c_s^2}{\tau} - \frac{\tau-\tau_0}{[1+(\tau-\tau_0)^2/\tau_B^2]\tau_B^2}\right) = 0 \ , \qquad (24)$$

Type-2： $$\partial_\tau \tilde{e} + (1+c_s^2)\left(\frac{\tilde{e}}{\tau} - b\frac{\tilde{e}}{\tau^2 T}\right) + \chi_m \sigma_0\left(\frac{2c_s^2}{\tau[1+(\tau-\tau_0)^2/\tau_B^2]^3} - \frac{3(\tau-\tau_0)\tau_B^6}{(\tau^2 - 2\tau\tau_0 + \tau_0^2 + \tau_B^2)^4}\right) = 0 \ , \qquad (25)$$

Type-3： $$\partial_\tau \tilde{e} + (1+c_s^2)\left(\frac{\tilde{e}}{\tau} - b\frac{\tilde{e}}{\tau^2 T}\right) + \chi_m \sigma_0 e^{-\frac{2|\tau-\tau_0|}{\tau_B}}\left(\frac{2c_s^2}{\tau} - \frac{1}{\tau_B}\frac{\partial|\tau-\tau_0|}{\partial_\tau}\right) = 0 \ , \qquad (26)$$

Figure 2 illustrates the effect of different magnetic field parameters on the evolution of energy density with proper time in the magnetized conformal fluid model, with the magnetic susceptibility $\chi_m$ fixed at 0.05. Figure 2 illustrates that stronger magnetic fields (higher $\sigma_0$ ) significantly slow the decay of energy density over proper time ($\tau$). For instance, the energy density for $\sigma_0$ = 25 (green dash-dotted line) decays much more slowly compared to weaker fields (e.g., $\sigma_0$ = 0.5, red short dashed line). This suppression arises due to magnetic pressure and Lorentz forces, which counteract the hydrodynamic expansion of the quark-gluon plasma (QGP). The magnetic field acts as a "brake" on energy dissipation by altering the fluid's equation of state and flow dynamics.

The three magnetic field evolution models in relativistic heavy-ion collisions exhibit distinct trends: Type-1 causes the strongest suppression of energy density decay. Type-2 shows rapid early-stage energy dissipation for weak fields ( $\sigma_0$ = 0.5), diverging from monotonic trends. Type-3 has the weakest overall impact on energy density evolution. These differences stem from the functional forms of the magnetic field decay (e.g., exponential vs. power-law dependence on $\tau$), which govern how magnetic energy is transferred to or stored in the fluid.

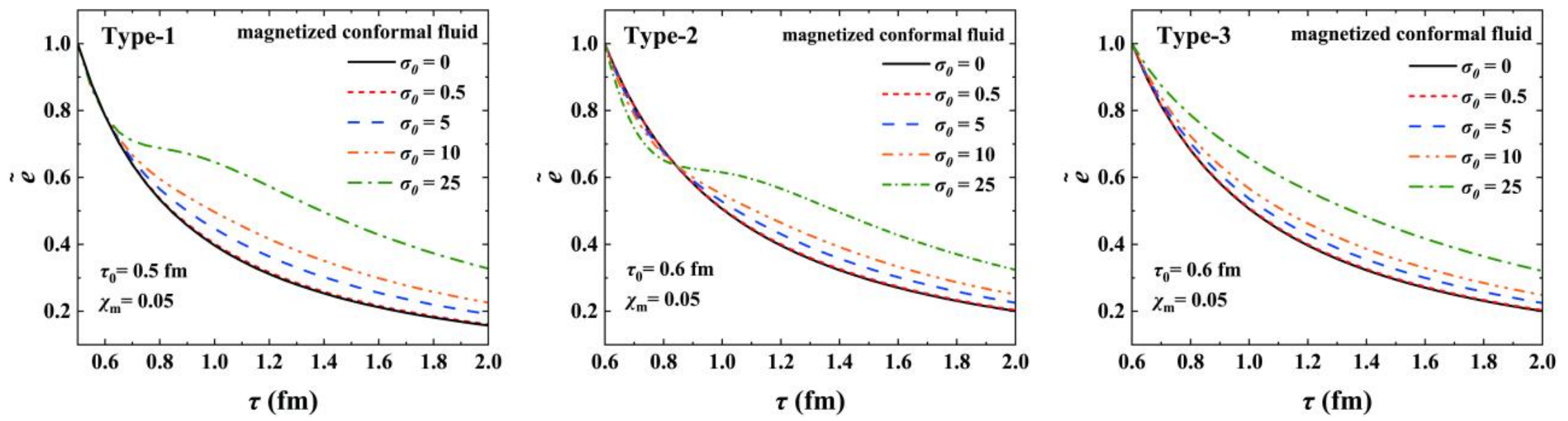

**Figure 2.** Evolution of energy density in the magnetized conformal fluid case. We choose $\chi_m = 0.05$. The left, middle, and right panels correspond to Type-1, Type-2, and Type-3 magnetic field time evolution, respectively. The black solid line represents the case without a magnetic field. The red short dashed line, blue long dashed line, orange dash-dot-dotted line, and green dash-dotted line correspond to $\sigma_0 = 0.5, 5, 10, 25$, respectively.

**Figure 3** compares the normalized energy density evolution $\tilde{e}(\tau)$ between an ultra-relativistic fluid (blue dashed line) and a magnetized conformal fluid (red dotted line) under identical initial conditions ( $\sigma_0$ = 25, $\chi_\mathrm{m}$ = 0.05). Please note that for a positive magnetization coefficient, the energy obtained by magnetized conformal fluids from the magnetic field is less than that obtained by ultra-relativistic fluids.

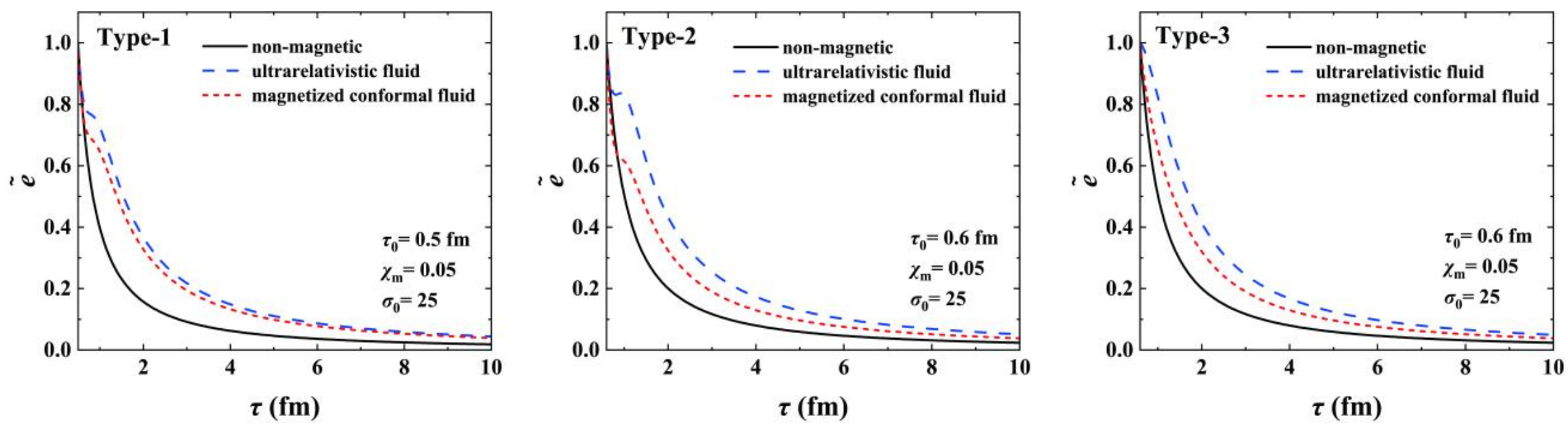

**Figure 3.** Evolution of the normalized energy density. The left, middle, and right panels correspond to Type-1, Type-2, and Type-3 magnetic field time evolution, respectively. The magnetic field parameter is set to $\sigma_0$ = 25, and the magnetic susceptibility is $\chi_m = 0.05$. The black line, the blue dashed line and red dotted line correspond to the non-magnetic fluid, ultra-relativistic fluid and the magnetized conformal fluid, respectively.

## V. Energy-density Evolution with Temperature-dependent Magnetic Susceptibility

Following the rather "physical" discussion in the previous section, we now consider a more realistic scenario based on lattice QCD. For simplicity, we choose the temperature $T$ as the independent variable and express the magnetic susceptibility $\chi_m$ as a function of $T$. In lattice QCD calculations, the behavior of magnetic susceptibility as a function of temperature has been investigated using several different approaches. The latest lattice QCD results on the magnetic susceptibility $\chi_m$ of QCD matter, which can be found in Ref.[60], show that at high temperatures, $\chi_m$ is parameterized in accordance with perturbative theory as follows

$$\chi_m\left(T\right) = 2\beta_1 \log(\frac{t}{q_0})\frac{1+g_0/t+g_1/t^2+g_2/t^3}{1+g_3/t+g_4/t^2+g_5/t^3}e^{\left(\frac{-h_3}{t}\right)}, \tag{27}$$

where $\beta_1 = 1/(6\pi^2)$ , $t$ parameter is defined as $T/1GeV$ ( $T$ is the temperature), which shown as a dimensionless quantity. These parameters $q_0$, $g_0, g_1, g_2, g_3, g_4, g_5$ and $h_3$ are 0.1544, 23.99, -2.085, 0.1290, 21.35, -6.201, 0.5766, 0.1497, respectively.

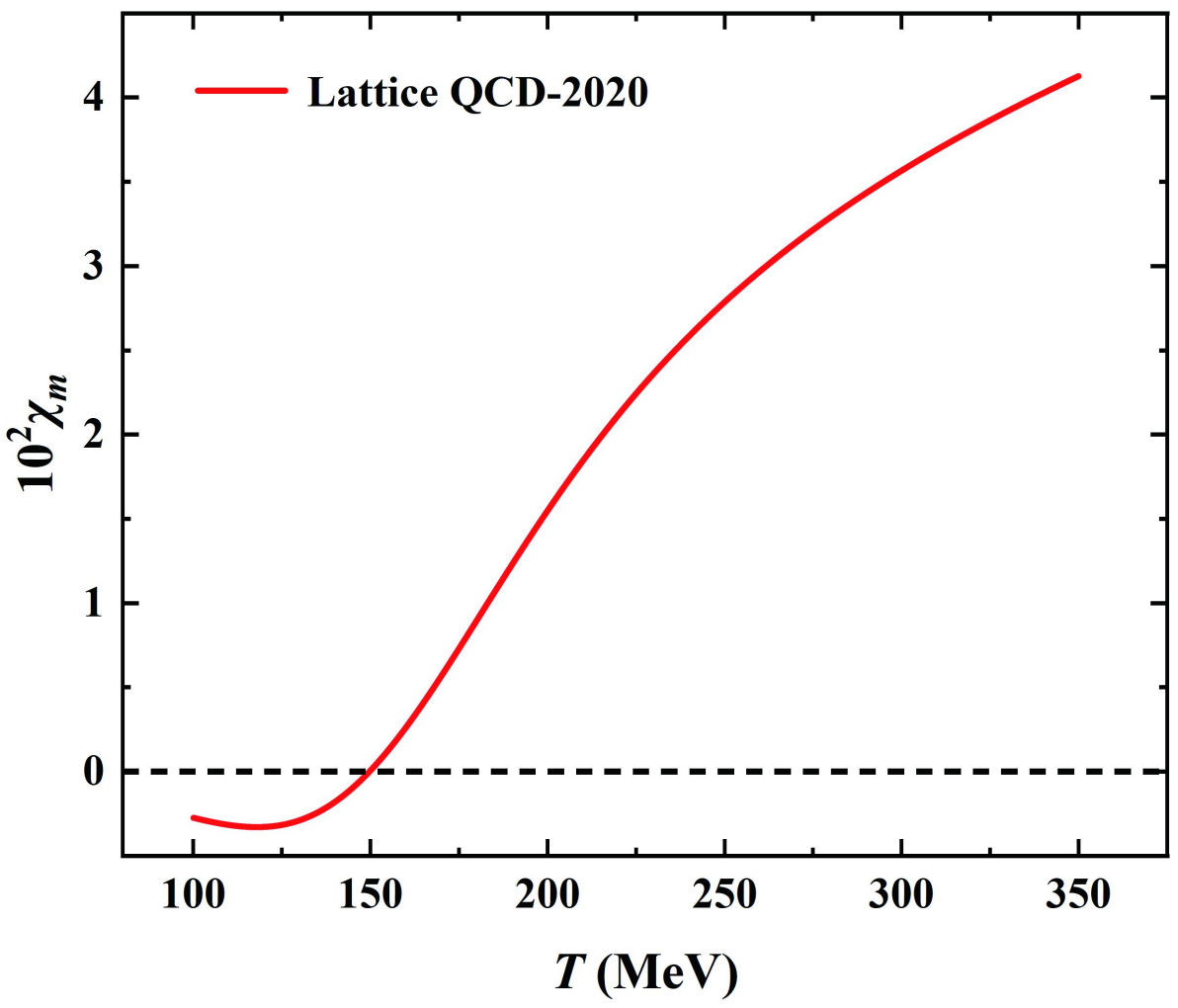

**Figure 4.** The magnetic susceptibility $\chi_m$ of QCD matter as a function of temperature. Results from lattice QCD calculations in 2020 [60].

Figure 4 presents the magnetic susceptibility $\chi_m$ of QCD matter as a function of temperature, derived from lattice QCD calculations conducted in 2020 [60]. The graph illustrates a significant increase in $\chi_m$ with rising temperature, transitioning from near-zero values at lower temperatures (approximately 100-150 MeV) to a steep ascent beyond 150 MeV. This trend suggests a marked change in the magnetic response of QCD matter as it transitions from hadronic to quark-gluon plasma (QGP) phases. In the confined phase, characterized by hadrons, the medium exhibits diamagnetic properties, with $\chi_m$, indicating a tendency to expel magnetic fields. Conversely, in the deconfined QGP phase, the medium becomes paramagnetic, with $\chi_m$, showing an increased magnetic response. Such temperature-dependent behavior is crucial for understanding the electromagnetic properties of QCD matter under extreme conditions, which is pertinent to the study of high-energy heavy-ion collisions and the evolution of early universe. The results from the lattice QCD-2020 calculations provide a valuable theoretical foundation for exploring the magnetic characteristics of QCD matter and their implications in high-energy physics, particularly in the context of phase transitions and the associated changes in magnetic susceptibility.

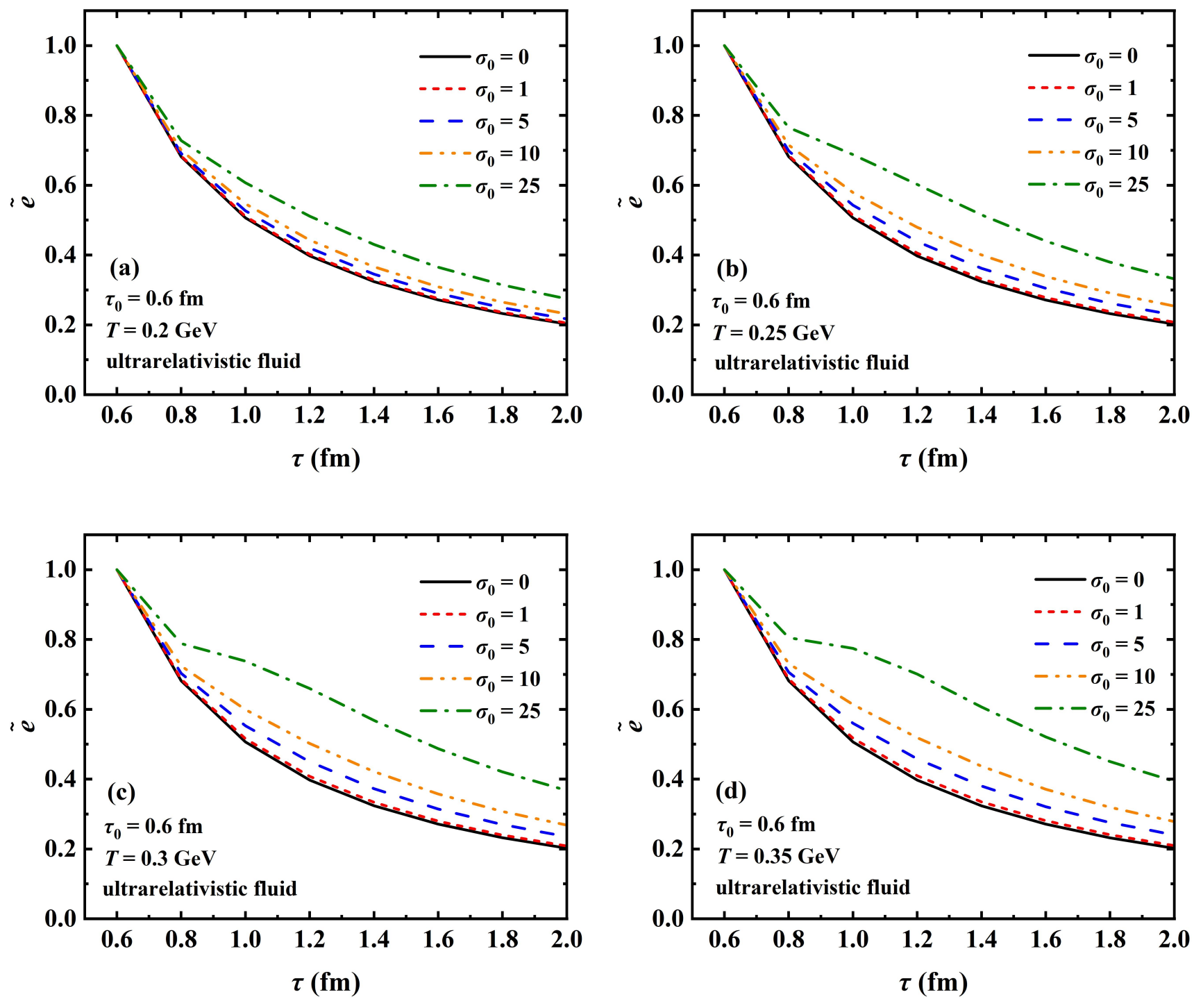

ments.

**Figure 5.** Temporal Evolution of the Magnetic Field in Type-2 Ultra-relativistic Fluids.Panels (a), (b), (c), and (d) correspond to different temperatures of the QGP deconfined phase ( $T$ = 0.2, 0.25, 0.3, 0.35 GeV), respectively. The black solid line represents the case without a magnetic field.

Figure 5 illustrates the temporal evolution of the magnetic field in Type-2 ultra-relativistic fluids. At $T = 0.2$ GeV, in the early stage after the confinement-deconfinement phase transition, where the magnetic susceptibility $\chi_m(T)$ as just started to rise from a near-zero value (paramagnetism emerges initially but remains weak), leading to a low coupling efficiency between the magnetic field and the fluid. The results demonstrate that temperature is the core regulating factor of the magnetic field-fluid coupling efficiency: as the temperature increases from 0.2 GeV to 0.35 GeV, the QGP magnetic susceptibility $\chi_m(T)$ is significantly enhanced, and the "sustaining effect" on energy density over the same magnetic field lifetime is improved synchronously.

Figure 6 illustrates the temporal evolution of the magnetic field in Type-2 magnetized conformal fluids. In this scenario, the energy density decay rate of the QGP is rather faster than that in ultra-relativistic fluids. This phenomenon stems from the synergistic effect between the equation of state and magnetic response of the magnetized conformal fluid: enhanced paramagnetism strengthens the coupling between the magnetic field and the fluid, yet it also induces more pronounced dissipation during the energy conversion between the magnetic field and the fluid. Combined with the ideal gas expansion characteristic of the QGP at elevated temperatures, this ultimately results in a steeper energy density decay slope than that of ultra-relativistic fluids under the same temperature and initial magnetic field strength $\sigma_0$ conditions. This finding reveals that the magnetization effect does not simply "enhance energy maintenance", but reshapes the energy evolution dynamics of the QGP through the "trade-off between coupling and dissipation". It clarifies the intrinsic difference in energy decay behavior between magnetized conformal fluids and ultra-relativistic fluids, providing a key observational signature for distinguishing fluid types in heavy-ion collision experiments.

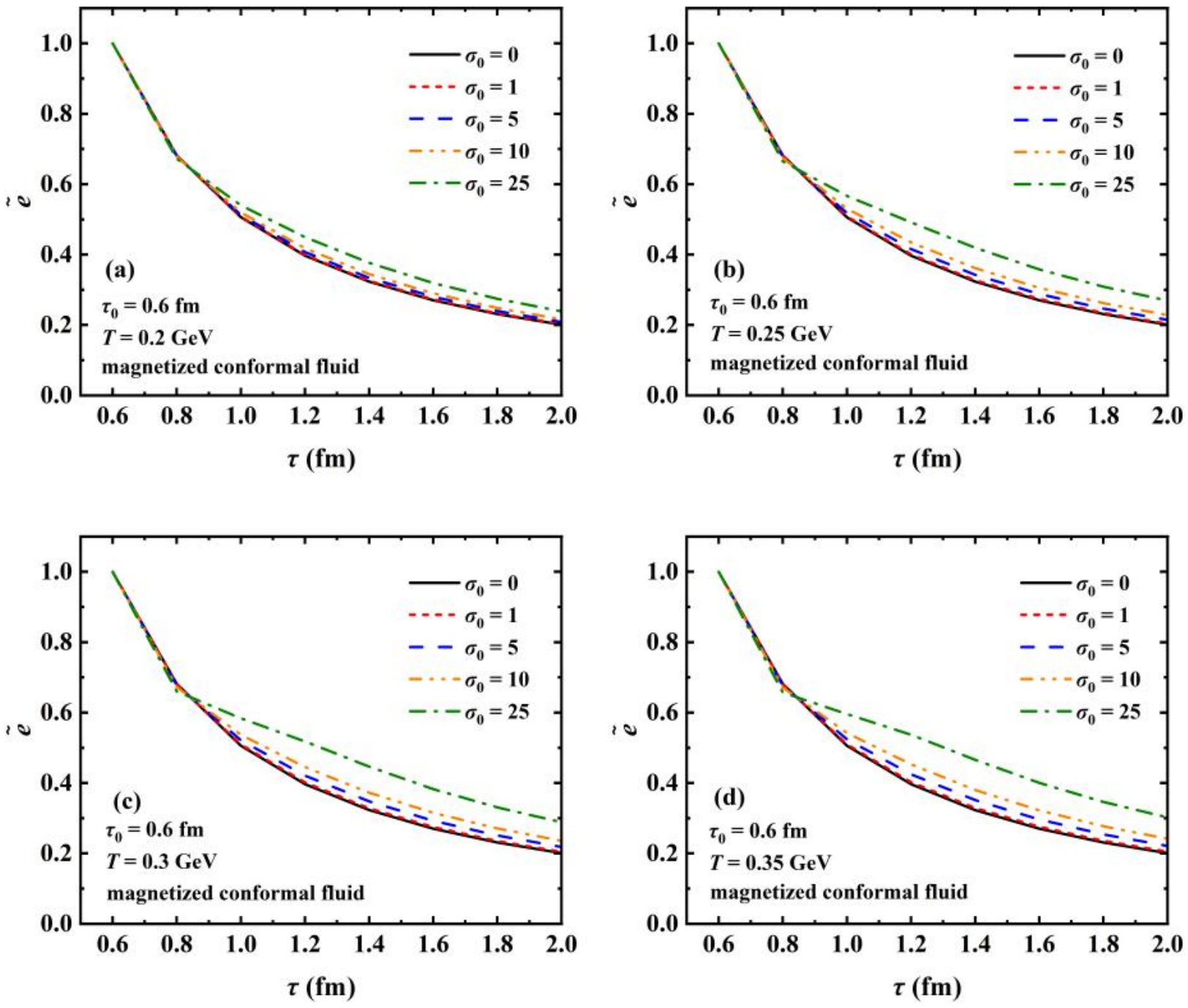

**Figure 6.** Temporal Evolution of the Magnetic Field in Type-2 Magnetized Conformal Fluids.Panels (a), (b), (c), and (d) correspond to different temperatures ( $T$ = 0.2, 0.25, 0.3, 0.35 GeV), respectively. The black solid line represents the case without a magnetic field.

The use of lattice QCD-derived $\chi_m(T)$ represents a significant step toward realistic modeling of magnetized QGP. It bridges theoretical predictions with experimental conditions in heavy-ion collisions and opens avenues for exploring the QCD phase diagram's electromagnetic signatures. By incorporating dynamic magnetic responses, this work paves the way for deeper insights into the interplay between quantum chromodynamics and relativistic magnetohydrodynamics.

## VI. Summary and Conclusions

In this work, we systematically investigated the evolution of energy density in relativistic fluids under time-dependent magnetic fields within the framework of relativistic magnetohydrodynamics (RMHD). By incorporating three distinct magnetic field evolution models (Type-1, Type-2, and Type-3) in relativistic heavy-ion collisions (RHIC) into the Bjorken flow framework, we analyzed their impacts on energy dissipation dynamics in both ultra-relativistic and magnetized conformal fluids. Furthermore, a temperature-dependent magnetic susceptibility derived from lattice QCD calculations was introduced to enhance the physical realism of the magnetic response in quark–gluon plasma (QGP). Our results demonstrate that magnetic fields play a critical role in modulating energy dissipation. Stronger magnetic fields consistently suppress the decay of energy density across all models, with the suppression magnitude being model-dependent. Notably, Type-1 exhibits the most pronounced retardation of energy dissipation, while Type-3 shows the weakest effect. This divergence arises from the distinct temporal evolution profiles of the magnetic fields, which govern how magnetic energy couples to the fluid. For instance, in ultra-relativistic fluids, the interplay between magnetic pressure and hydrodynamic expansion leads to slower thermalization, while in magnetized conformal fluids, explicit magnetization terms in the equation of state amplify energy retention.

It is worth noting that the magnetic suppression of energy density decay observed here shares qualitative similarities with the effects of bulk viscosity in Bjorken flow [Ref. [61], Figure 3]. Both mechanisms introduce effective pressure contributions that counteract hydrodynamic expansion. In realistic heavy-ion collisions, the actual evolution likely reflects a combination of magnetic and viscous effects.

The inclusion of temperature-dependent magnetic susceptibility further enriches the dynamics. Lattice QCD-based parametrizations reveal a transition from diamagnetic be havior in the confined phase to paramagnetic response in the deconfined QGP phase. This temperature dependence introduces a feedback mechanism: delayed energy dissipation sustains higher temperatures, reinforcing paramagnetic behavior and influencing the magnetic field's evolution. Such findings underscore the necessity of integrating QCD-based susceptibilities into RMHD frameworks for accurate modeling of heavy-ion collisions.

This study bridges theoretical RMHD models with experimental observables, emphasizing the importance of precise magnetic field modeling in interpreting QGP dynamics. However, several limitations warrant future exploration. First, dissipative effects such as viscosity [56,62] and finite conductivity—neglected in the ideal RMHD approximation— could modify energy-momentum transfer. In addition, future work could explore the attractor nature of these solutions by identifying initial conditions that lead to identical asymptotic behaviors, as discussed in studies of viscous hydrodynamics.